\documentclass[superscriptaddress,
twocolumn,
showkeys, preprintnumbers,
floatfix,]{revtex4-2}

\usepackage{dynlearn}

\theoremstyle{definition}

\newcommand{\ourTitle}{
Whale Casting: \\
Remote mobile streaming \\
humpback whale vocalizations\\
to the world
}

\begin{document}

\title{\ourTitle}

\author{\href{http://csc.ucdavis.edu/~chaos}{James P. Crutchfield}}
\email{chaos@ucdavis.edu}
\affiliation{\href{http://csc.ucdavis.edu}{Complexity Sciences Center},
Physics and Astronomy Department\\
University of California, Davis, California 95616}

\author{\href{http://csc.ucdavis.edu/~ajurgens/}{Alexandra M. Jurgens}}
\email{amjurgens@ucdavis.edu}
\affiliation{\href{http://csc.ucdavis.edu}{Complexity Sciences Center},
Physics and Astronomy Department\\
University of California, Davis, California 95616}

\date{\today}
\bibliographystyle{unsrt}

\begin{abstract}
Over several days in early August 2021, while at sea in Chatham Strait,
Southeast Alaska, aboard M/Y Blue Pearl, an online
\href{https://www.twitch.tv/https://www.twitch.tv/}{twitch.tv} stream broadcast
in real-time humpback whale vocalizations monitored via hydrophone. Dozens on
mainland North American and around the planet listened in and chatted via the
stream. The webcasts demonstrated a proof-of-concept: only relatively
inexpensive commercial-off-the-shelf equipment is required for remote mobile
streaming at sea. These notes document what was required and make
recommendations for higher-quality and larger-scale deployments. One conclusion
is that real-time, automated audio documenting whale acoustic behavior is
readily accessible and, using the cloud, it can be directly integrated into
behavioral databases---information sources that now often focus exclusively on
nonreal-time visual-sighting narrative reports and photography.
\end{abstract}


\preprint{arxiv.org:2212.XXXXX}

\maketitle

\section{Introduction}
\label{sec:introduction}

Humpback whales (\emph{Megaptera Novaeangliae}) are well-known for their
extensive vocalizations, with dozens of short social calls believed to be used
for group coordination and long, ``melodic'' songs from males believed to play
a role in mate selection and competition \cite{Payn71b}. The first broad
appreciation of their vocal repertoire stemmed from hydrophone recordings made
in the 1960s---captured on Roger Payne's 1970 LP record
\href{https://www.amazon.com/Songs-Humpback-Whale-Roger-Payne/dp/B000UTBPTG/ref=sr_1_3}{\emph{Songs
of the Humpback Whale}} \cite{Payn70a}. The LP sold 100,000s of copies and,
eventually through increased human awareness, aided in the passage of
international bans on whale hunting by the late 1970s.

Humpback whale ancestors---in the order \emph{cetacea}---appeared in the
planet's oceans some 56 Myrs ago, having evolved from ungulate land
mammals---the ancestors of the modern-day hippopotamus. Their long
evolution---exceeding that of human's by an order of magnitude---led to complex
social organizations, tool-use (socially-coordinated bubble-net feeding), and
hemisphere-spanning acoustic communication \cite{Payn71a}.

Though they must surface to breathe, their immediate experience largely occurs
submerged. During their dives to feed and socially interact, ambient light
rapidly dims at depths greater than $50$ m. As a result, their
world-experience---their \emph{umwelt}---is predominantly acoustic. And, given
the preponderance of time under the sea surface, out of sight from observers,
they must be studied via hydrophone monitoring of their underwater
vocalizations. The resulting technical and logistical challenges add to the
mystery of their lives.

Much evidence, if anecdotal, has accumulated that points to their native
intelligence. They exhibit advanced intentional behaviors and conscious
awareness through their raw intelligence, song generation and sharing
\cite{Payn83a,Garl13a,Garl17a}, communication and interactions with their own
and other species \cite{Smit08a,Chol18a}, and empathy (concern for other's
well-being) \cite{Pitm17a}.

How does such cognition manifest in humpback whale communicative acoustic
interactions? Success in addressing this will both substantially enhance their
conservation and advance our appreciation of co-existing, independent
intelligent animals on the planet. The research challenge is daunting, however,
as the principal observations come from hydrophone recordings occasionally
supplemented with surface visual and acoustic observations made from ocean
vessels in remote locations, sometimes far at sea. The animals are
too large for study in captivity---a mode of research now deprecated
for even smaller marine mammals.

Technological advances promise much improved understanding, though. Recently,
for example, digital video-sound recording tags are being actively
attached to swimming whales, providing acoustic, visual, and environmental data
that can reveal new aspects of their undersea behaviors over tens of minutes to
hours \cite{Gold17a}. Recent, handsomely-funded efforts promise to deploy such
technologies on a wholly new level of large-scale multi-modal monitoring and
automated detection and data analysis \cite{Andr22a}.

One goal in all this is to infer the meaning content of humpback vocalizations.
However the data is obtained, one approach to probe meaning is to correlate
recorded vocalizations with observed social behavior. What kinds of social
interaction might provide workable contexts in which to extract the semantics of
vocalizations?

Humpbacks, especially in Southeast Alaska, are notable for their
tool-use---socially-coordinated bubble-net feeding. In relatively small groups
(half dozen to a dozen) they coordinate to improve their feeding. And, this is
done, it appears, through much vocalization between group members as several in
the group construct a cylindrical curtain of bubbles that localizes the prey
(krill or herring). This culminates in a group member vocalizing a ``feeding
call'' that initiates the lunge feeding of conspecifics waiting below who swim
to the surface with mouths wide open to engulf a column of food. In this,
humpbacks are an excellent study animal for correlating behavior and the
function of social calls and so for extracting call semantics.

What experimental design and protocols support such studies? The challenge is
to collect sufficient data of the animals' trajectories, vocalizations, and
visual behavioral observations so that vocalizations can be placed within a
context of functional behavior. To afford the appropriate statistical error
analysis, the key here is sufficiency: long-term longitudinal data
automatically collected over a spatial range large enough to appropriately
cover humpback territory. The result would be an extremely large database.
Initially, this would likely tax modern capabilities. Current trends,
though, indicate that within several years automated analysis would be quite
workable.

One overall strategy to achieve this is to deploy a whale ``observatory'' in
grounds that are regularly visited. The observatory would consist of extensive
sound recording and sound generating transducers along with autonomous airborne
and undersea drones to visually monitor behaviors. The data flow to and from
the transducers would be transmitted via high-speed Wifi radio links to
land-based base-stations. The latter would then be connected via satellite link
to the Internet.

This vision led to the 2019 proposal for SEAWHO---the SouthEast Alaska WHale
Observatory; cf. the presentation
\href{http://csc.ucdavis.edu/~chaos/papers/WhaleOverviewTrim.mp4}{N Whales from
M Hydrophones}. It should be noted that SEAWHO is, at present, modest compared
to more recent efforts now ramping up. For example, there is the large and
well-funded Project CETI that will study sperm whales in the Atlantic
\cite{Andr22a}. There is also interest in whale sanctuaries that provide safe
environments for previously-captive marine mammals. The
\href{https://whalesanctuaryproject.org/}{Whale Sanctuary Project} comes
immediately to mind.

The following addresses only a part of the technology SEAWHO requires: Internet
streaming of humpback vocal behaviors from remote locations at sea. It starts
describing an August 2021 voyage that implemented a very low-cost system. It
then reviews issues related to the physics of the whale's water world and sound
propagation. It recounts the technical details of live streaming and doing so
remotely while at sea. Finally, it concludes listing resources and proposing
directions for future efforts that take advantage of rapidly-advancing
technologies to move closer to SEAWHO.

Note that real-time undersea sound is already available on the web. For some
time, as an example, the Monterrey Bay Aquarium Research Institute has supported
a \href{https://www.mbari.org/soundscape-listening-room/}{live link} to
acoustic signals deep in Monterrey Bay, California, detected by an anchored
hydrophone. The following report also concerns a live ocean-acoustic link, but
with complementary motivations: the system is mobile, easily portable, deployed
remotely, and inexpensive. This makes acoustic monitoring of whale vocalization
widely available. In the context of open science, this could greatly accelerate
our learning much more (and more quickly) about the lives of whales; at least,
those that are vocal.

\begin{figure*}[t]
\centering
\includegraphics[width=.98\textwidth]{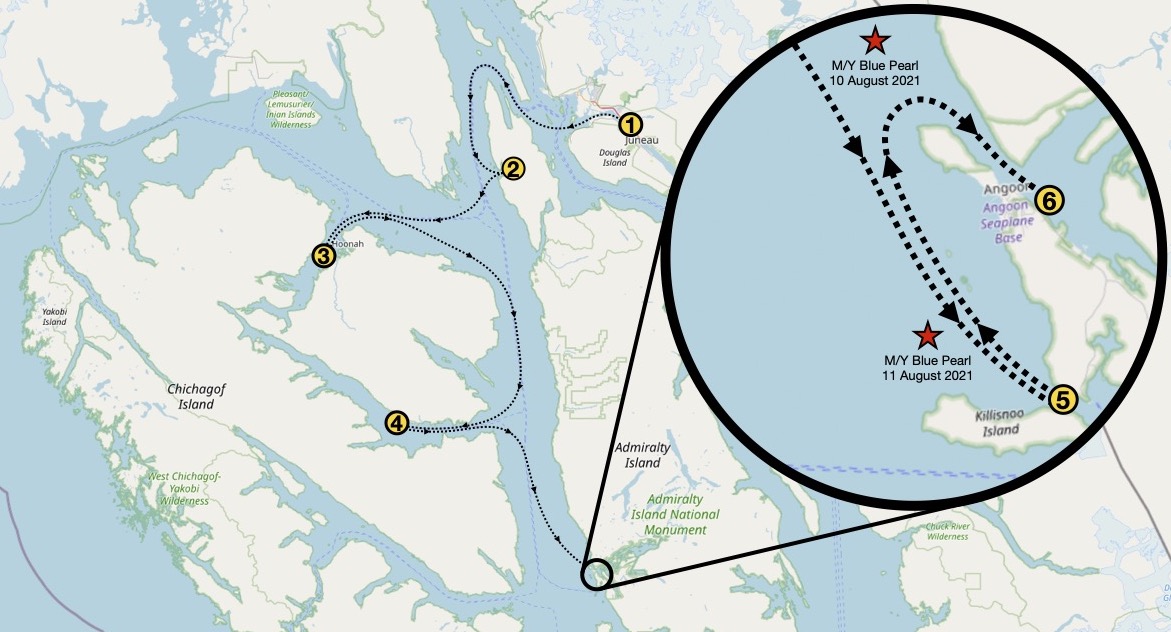}
\caption{Streaming humpback whale vocalizations in Chatham Strait, anchored
	off Angoon, Alaska. The red stars denote the location of M/Y Blue Pearl on
	the two successive streaming days, 10 and 11 August 2021.
	\href{https://www.openstreetmap.org/copyright}{OpenStreetMap}
	\href{https://creativecommons.org/licenses/by-sa/2.0/}{CC BY-SA 2.0}.
    }
\label{fig:AngoonTwitchStream}
\end{figure*}

\begin{figure*}[t]
\centering
\includegraphics[width=.95\linewidth]{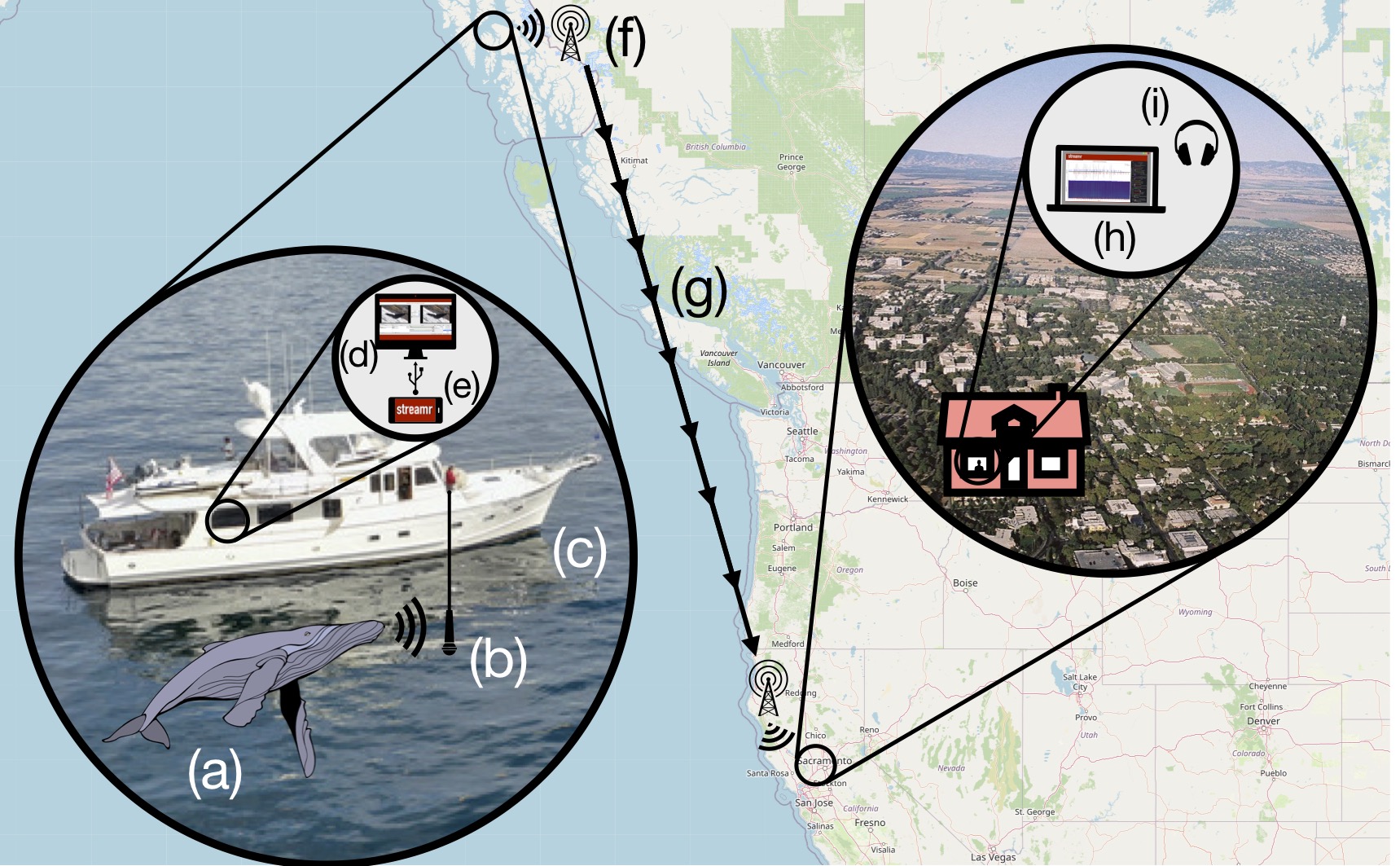}
\caption[text]{Remote mobile whale streaming: Humpback whales (\emph{Megaptera
	Novaeangliae}) (a) vocalizing and hydrophone (CT54, Cetacean Research
	Technologies) (b) submerged $15'-30'$ beneath vessel (M/Y Blue
	Pearl, Vancouver, Canada) (c) converts underwater acoustic waves to
	electronic signal linked via secured cable to MacBook Pro 15" computer (d)
	on the Blue Pearl's flying bridge that recorded the hydrophone signal. The
	computer (d) was tethered to an iPhone 5s smartphone (e) connected
	line-of-sight to a land-based cell tower (f) on the Internet (g). In this
	way, the system webcast the humpback vocalization twitch stream through the
	cloud to a laptop computer (h) somewhere in the world that is being
	listened to on headphones (i).
    }
\label{fig:WebcastConfiguration}
\end{figure*}

\section{Voyage}

The voyage departed Aurora Harbor, Juneau, Alaska, on 6 August 2021, on M/Y
Blue Pearl, a \href{https://www.flemingyachts.com/model/Fleming-65}{Fleming 65}
raised pilothouse motoryacht captained by Don and Denise Bermant. The voyage
headed northwest to meet Chatham Strait and then headed south, anchoring in
Funter Bay, Admiralty Island, (6 August) and at Hoonah (7 August) and Tenakee
Springs (8 August) on Chicagoff Island. Few humpback whales were seen. From
that point forward, however, motoring south through Chatham Strait the M/Y Blue
Pearl encountered many, often performing their well-known socially-coordinated
bubble-net feeding.

Due to the high number of whales actively feeding, the voyage spent several
days on the western coast of Admiralty Island, first anchored at Killisnoo
Island and then docked in Angoon Harbor. Day trips were taken out into Chatham
Strait to observe the humpbacks, documenting their behaviors visually via
vessel photography and aerial photography (drone), along with acoustically
monitoring via hydrophone.

Vocalizations from the whale groups were particularly notable during bubble-net
feeding: many minutes of a cacophony of diverse animal calls, apparently from
many individuals, culminating in a distinctive frequency-upsweep ``feeding
call'', seemingly from a single individual. Within just a minute or two of that
call the feeding whales, that had been waiting below the bubble-net, breached
the surface, mouths wide open to engulf the prey. Taking advantage of this
activity, the voyage cruised the waters off Angoon over several days:
Tuesday-Wednesday 9-11 August.

Having set up and tested the computing and recording equipment and software
(described shortly), we streamed the underwater sounds picked up by hydrophone
on August 10th and on August 11th. See Fig. \ref{fig:AngoonTwitchStream} for
the locations, at which the smartphone had line-of-sight connection to cell
towers in the town of Angoon. Adapting to variable cell-signal quality, we
streamed for several hours each day using
\href{https://www.twitch.tv/}{twitch.tv} on channel
\href{https://www.twitch.tv/drjpchaos}{DrJPChaos}.

\section{Ocean Acoustiphysics}

Sound propagation in water differs markedly and in key ways from propagation in air. Given human's innate sense and experience of sound in air, the differences need to be taken into account when interpreting the signals that hydrophones pick up.

First, the speed of sound in water is five times that in air: $1,500$ meters
per second compared to $340$ meters per second, respectively, owing to the
water medium being markedly denser than air. Practically, this leads to, for
example, echos as sounds bounce off the seabed. Since density increases with
depth, water depth is important and, of course, changes when changing
anchorages. This also means that sounds from distant sources can be detected.
For example, one is often surprised by the degree to which vessel noise is
heard and in some cases dominates the undersea soundscape, even if vessels are
not in sight. Commercial cruise liners are notable contributors to ocean noise
given their immense displacement (key to waves generated by their passing) and
massive engines.

\begin{figure*}[t]
\centering
\includegraphics[width=.98\linewidth]{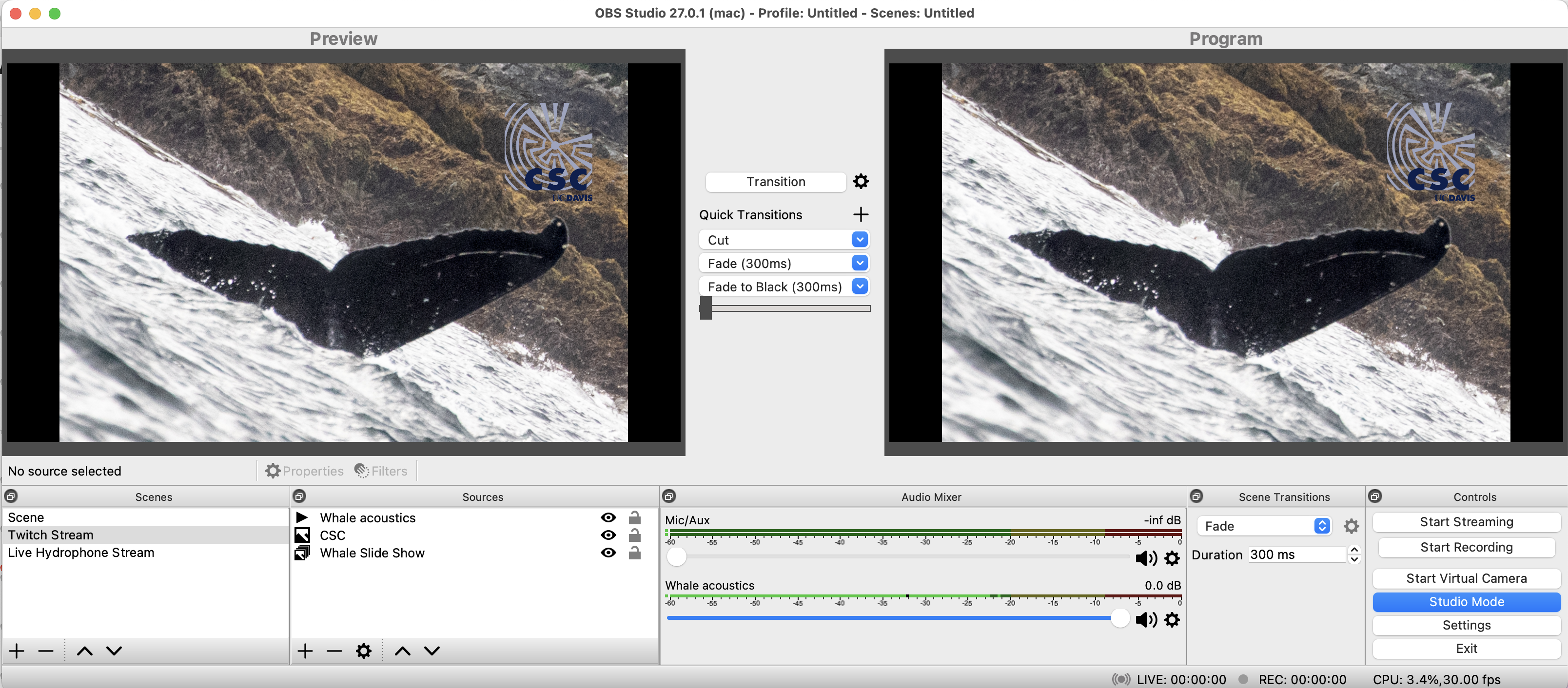}
\caption[text]{OBS Studio: Screenshot of running the twitch stream on board
	M/Y Blue Pearl anchored off Angoon, Alaska. The left window (Preview) allows
	the user to see and compose the stream components: live or recorded video,
	labels, images, and the like. Meanwhile, the live stream appears in the
	right window (Program). Below the two windows, the audio mixer shows the
	current status of one or several audio clips, either prerecorded or
	real-time. The available stream components are listed to the left.
	Due to limited bandwidth for streaming, only static images were used as
	visual background, not live video.
	}
\label{fig:OBSScreenshot}
\end{figure*}

Second, the precise nature of propagation in water is complicated by the fact
that sound velocity increases with water pressure (and so depth) and decreases
with water temperature and salinity.

Third, unlike sound in air, underwater sound at different frequencies
propagates at different speeds---this is referred to as frequency dispersion.
Thus, a distinct sound pulse detected at some distance loses its sharpness and
blurs out over a time period much longer than the original pulse.

Taken altogether, the effects of these dependencies have on propagation are
unlike those of our experience of sound in air. They often result in unusual
and counterintuitive sound phenomena. The physics underlying these effects are
nicely recounted in Ref. \cite{Payn95a}.

For example, the dependencies lead to a fascinating phenomenon of extremely
long-ranged detection of sound signals in the ocean. This is the \emph{Sofar
channel}. Due to the competing effects of pressure and temperature on sound
speed, there is a horizontal ``channel'' that conducts sounds like a waveguide:
signals within a certain frequency band bounce between a shallow ``ceiling''
(perhaps $10$s of meters in depth) and a ``floor'' ($100$s meters or more in
depth). The net result is that sound signals in the Sofar channel can propagate
very long distances---easily tens of kilometers or, depending on conditions, to
hundreds or thousands of kilometers.

Given the undersea is their environment and given their evolution over millions
of years, whales have accounted for and take advantage of these ocean-acoustic
properties. These features affect what they can perceive, how they generate
sound underwater, and how they communicate and socialize. Undoubtedly, many
aspects of their vocalizations are naturally adapted.

Finally, these properties affect the acoustic signals one records via a
hydrophone and so, too, how one interprets what one is hearing.

\section{Design for Whalecasting}
\label{sec:Design}

To capture humpback vocalizations in real-time we deployed a ``dip'' hydrophone
(Model C54, \href{https://www.cetaceanresearch.com}{Cetacean Research
Technology}, Seattle, WA) suspended at a depth of approximately $5-10$ m
($15'-30'$) below the vessel. On occasion, though, strong local currents varied
this depth considerably due to the wind-blown vessel dragging the hydrophone.
Note that the currents in Southeast Alaska are powerful---largely driven by
substantial $6$ m ($15'-20'$) tides and constrained by the complex seabed
and island shoreline topography.

Via physically-secured, long audio cables, the hydrophone signal was recorded
by a laptop (Macintosh MacBook Pro 15'') on the M/Y Blue Pearl's flying bridge.
This above-water-line vantage point was extremely helpful in sighting and
following whales. For vessel and equipment safety, the hydrophone was never
deployed while the vessel's engine was operating. Recording sample rates were
set at $44.1$ kHz with $16$ bits per sample.

To prepare, edit, and monitor audio files we used both
\href{https://www.audacityteam.org}{Audacity} (v. 3.0.2) and
\href{https://ravensoundsoftware.com/software/raven-pro/}{Raven Pro} (v. 1.6).
They provide waveform and spectrogram views of audio signals, but each
has complementary user interfaces that are convenient in different real-time
operation settings.

Figure \ref{fig:WebcastConfiguration} presents a schematic of the overall
setup.. High quality and sound-isolating headphones were essential, given
ambient sounds and often subtle ocean sounds coming in from the hydrophone.

For video recording and live streaming we used open source software \href{https://obsproject.com/}{Open Broadcast Software OBS Studio 27.0.1}.
It merges images and audio and video signals and makes the connection to the
twitch streaming engine in the cloud. Figure \ref{fig:OBSScreenshot} presents a
screenshot of OBS Studio operating on the laptop.

\section{Cloud access}

A medium- to high-speed link to the Internet was essential for access to
twitch.com and smooth streaming. This was helped in large measure by finding a
line-of-sight connection to a land-based cell tower in Angoon. This limited the
flexibility of choosing our anchorages. This combined with the humpback's rapid
movements---at the surface they often cruise around 10 kph---complicated
planning and determining when to webcast.

There is good online documentation for configuring streaming parameters and
using twitch itself and for setting up and running OBS Studio
(\href{https://obsproject.com/wiki/OBS-Studio-Quickstart}{OBS Studio
Quickstart}). Needless to say, configuring and testing the software and online
performance were extensively explored prior to the voyage. This was key to
establishing a number features necessary for smooth streaming and for minimizing
dropping and restarting the connection.

For the functionality reported here the free version of twitch.com was adequate;
that is, no subscription was required. The initial impression of twitch,
though, is dominated by its strong push to monetize one's channel.
Fortunately, this and its heavy merchandising can be avoided once a channel is
configured and running. Twitch's content has a bias towards gaming and a very
young demographic.

The DrJPChaos channel was featured on our research websites:
\href{http://VoicesOfTheDeep.org}{Voices of the Deep} and
\href{http://WorldWideWhale.org}{World Wide Whale}. The
\href{http://demon.csc.ucdavis.edu/~chaos/share/www/WorldWideWhale/WhaleCast/WhaleCast.html}{whale
casting page} is hosted on the latter.

\section{Conclusion}

Times and technology have changed immeasurably since Roger Payne's LP
recording. The challenge now is how to actively shape our future understanding
of whale communication in the wild. The results suggest a coming era of citizen
marine social science. Whale casting---remote mobile streaming of whale
vocalizations---gives a practical and relatively inexpensive path to it.

This report outlined a very modest, but accessible implementation using
inexpensive commercial-of-the-shelf (COTS) hardware and software. The net
functionality allowed for real-time streaming of humpback whale vocalizations
from remote locations on a mobile at-sea platform. The main constraint, as
noted, was the need to find in these locations line-of-sight connections to
cell towers. Nonetheless, the report outlined a path that moves us closer to
SEAWHO and improved appreciation of the lives of humpback whales.

The successful proof-of-concept suggests substantial improvements. So, one can
look forward to future implementations that provide higher quality and
multichannel sound and video streaming and larger-scale implementations that
support multiple hydrophones and video monitoring and real-time monitoring,
recording, and analysis in the cloud.

Perhaps most critical to achieving these is the recent emergence of high-speed
mobile links to satellite Internet. For example,
\href{http://www.starlink.com/rv}{In-Motion Starlink} recently arrived to
support mobile webcasting, which requires high-speed, low-latency internet
access. Notably, this just became an option for the tests recounted here as southeast Alaska become an area rated for ``high capacity'' coverage.

Specifically, Flat High Performance Starlink promises to provide high-speed,
low-latency internet communication while in-motion. Given the new system's wide
field of view and enhanced GPS, it connects to broader range of satellites for
consistent connectivity while moving. There are as yet no reports of shipboard
deployments, though. Given the antenna's design, deployments will likely
require calm seas or, more elaborately, motion compensation to stabilize
antenna and so reception.

\section*{Acknowledgments}
\label{sec:acknowledgments}
\vspace{-0.2in}

The authors are especially grateful to Captain Don Bermant and Denise Bermant
for their generous circumnavigation of Admiralty Island and environs aboard the
M/Y Blue Pearl. The voyage around Admiralty Island was arranged with them by
Tony Gilbert of The International Sea Keepers Society (seakeepers.org). The
authors thank Don Bermant, Denise Bermant, Jodi Frediani, Tony Gilbert,
Brenda McCowan, and Fred Sharpe for helpful discussions during the voyage.
JPC thanks Fred Sharpe for the use of a Cetacean Research Technologies model
C54 hydrophone. The effort was conducted under his National Marine Fisheries
Service Research Permit \#19703.

{\bf Author contributions}:
J.P.C. designed and conducted the experiments using personal and
University of California equipment. Both authors participated in all aspects of
manuscript writing and production.

{\bf Funding}: The authors' efforts were supported by, or in part by, Templeton
World Charity Foundation Diverse Intelligences grant TWCF0570 (Lead P.I. J. P.
Crutchfield) and Foundational Questions Institute and Fetzer Franklin Fund
grant FQXI-RFP-CPW-2007 (Lead P.I. J. P. Crutchfield) to the University of
California, Davis, and TWCF grant TWCF0440 to the SETI Institute (Lead P.I. L.
Doyle; Co-PIs J. P. Crutchfield, M. Fournet, B. McCowan, and F. Sharpe). The
opinions expressed in this report are the authors' and do not necessarily
reflect the views of Templeton World Charity Foundation, Inc.

{\bf Competing Interests}: None declared.

{\bf Data and materials availability}: Data provided by the first author upon
reasonable request.

\end{document}